\documentclass[article, 10pt, twocolumn, twoside]{IEEEtran}
\usepackage{theorem}
\usepackage{amsmath}
\usepackage{amssymb}
\usepackage{graphicx}
\usepackage{caption}
\usepackage{enumitem}
\usepackage{cite}
\usepackage[english]{babel}
\usepackage{epsfig,amssymb,amsfonts}
\usepackage[latin1]{inputenc}%latin1
\usepackage{amsmath}
\usepackage{amssymb}
\usepackage{cite}
\usepackage{bbm}
\usepackage{subfigure}
\usepackage{tikz}
\usepackage{tkz-berge}
\usepackage{pgfplots}
\usetikzlibrary{plotmarks}
\usepackage{multicol,float}
\usetikzlibrary{positioning}

\def\Ab{\mathbf{A}}
\def\Bb{\mathbf{B}}
\def\Cb{\mathbf{C}}
\def\Db{\mathbf{D}}
\def\Eb{\mathbf{E}}
\def\Fb{\mathbf{F}}

\def\Iden{\mathbf{I}}

\def\Kb{\mathbf{K}}
\def\Mb{\mathbf{M}}

\def\Pb{\mathbf{P}}
\def\Qb{\mathbf{Q}}
\def\Rb{\mathbf{R}}
\def\Sb{\mathbf{S}}
\def\Tb{\mathbf{T}}
\def\Vb{\mathbf{V}}
\def\Ub{\mathbf{U}}
\def\Wb{\mathbf{W}}

\def\eb{\mathbf{e}}
\def\ub{\mathbf{u}}
\def\wb{\mathbf{w}}

\def\pb{\mathbf{p}}

\def\gb{\mathbf{g}}

\def\yb{\mathbf{y}}
\def\xb{\mathbf{x}}
\def\zb{\mathbf{z}}

\def\Sig{\boldsymbol{\Sigma}}

\def\Lamb{\boldsymbol{\Lambda}}
\def\lamb{\boldsymbol{\lambda}}
\def\Pib{\boldsymbol{\Pi}}

\def\mub{\boldsymbol{\mu}}
\def\Phb{\boldsymbol{\Phi}}
\def\phb{\boldsymbol{\phi}}
\def\Xib{\boldsymbol{\Xi}}
\def\thb{\boldsymbol{\theta}}
\def\0b{\boldsymbol{0}}

\def\cg{\left[}
\def\cd{\right]}
\def\ag{\left\{}
\def\ad{\right\}}
\def\pg{\left(}
\def\pd{\right)}

\def\vec{\textup{vec}}
\def\resp{\textup{resp. }}

\def\E{\mathbb{E}}

\newtheorem{Thm}{Theorem}[section]
\newtheorem{remark}{Remark}[section]

\newtheorem{lemma}{Lemma}[section]

\title{On the asymptotics of Maronna's robust PCA}
\author{Gordana Dra\v{s}kovi\'{c},~\IEEEmembership{Student Member,~IEEE}, Arnaud Breloy,~\IEEEmembership{Member,~IEEE} and Fr\'{e}d\'{e}ric Pascal,~\IEEEmembership{Senior Member,~IEEE}\thanks{Gordana Dra\v{s}kovi\'{c} and Fr\'{e}d\'{e}ric Pascal are with L2S - CentraleSup{\'e}lec - CNRS - Universit\'{e} Paris-Sud - 3 rue Joliot-Curie, F-91192 Gif-sur-Yvette Cedex, France (e-mails: gordana.draskovic@l2s.centralesupelec.fr, frederic.pascal@l2s.centralesupelec.fr), Arnaud Breloy is with LEME - EA4416, University Paris-Nanterrre, France (e-mail: abreloy@parisnanterre.fr).}}

\begin{document}

\maketitle

\begin{abstract}
The eigenvalue decomposition (EVD) parameters of the second order statistics are ubiquitous in statistical analysis and signal processing. Notably, the EVD of robust scatter $M$-estimators is a popular choice to perform robust probabilistic PCA or other dimension reduction related applications. Towards the goal of characterizing the behavior of these quantities, this paper proposes new asymptotics for the EVD parameters (i.e. eigenvalues, eigenvectors and principal subspace) of the scatter $M$-estimator in the context of complex elliptically symmetric distributions. First, their Gaussian asymptotic distribution is obtained by extending standard results on the sample covariance matrix in a Gaussian context. Second, their convergence rate towards the EVD parameters of a Gaussian-Core Wishart Equivalent is derived. This second result represents the main contribution in the sense that it quantifies when it is acceptable to directly plug-in well-established results on the EVD of Wishart-distributed matrix for characterizing the EVD of $M$-estimators. Eventually, some examples (low-rank adaptive filtering and Intrinsic bias analysis) are provided to illustrate where the obtained results can be leveraged.
\end{abstract}

\section{Introduction}
The second order statistics plays a key role in signal processing and machine learning applications. Usually, this parameter is unknown and has to be estimated in order to apply a so-called adaptive process. In this scope, the $M$-estimators of the scatter \cite{Maronna76, tyler1987distribution} have attracted a lot of interest \cite{gini2002covariance, pascal2008covariance, ollila2012complex ,ollila2014regularized, sun2014regularized, Wiesel2015book} due to their robustness properties over the family of Complex Elliptically Symmetric (CES) distributions \cite{ollila2012complex}. They notably offer robustness to outliers and heavy tailed samples (now common in modern datasets), where the traditional Sample Covariance Matrix (SCM) usually fails to provide an accurate estimation.

The statistical characterization of the $M$-estimators of the scatter is a complex issue since they are defined by fixed-point equations. While the SCM in a Gaussian setting follows a well-known Wishart distribution \cite{muirhead1982aspects}, the true distribution of the $M$-estimators remains unknown. 
Several works derived asymptotic characterization for these estimators. 
Their asymptotic Gaussian distribution was derived in \cite{tyler1982radial} and extended to the complex case in \cite{ollila2012complex, mahot2013asymptotic}. 
Probably approximately correct (PAC) error bounds have been studied in \cite{soloveychik2015}.
Their analysis in the large random matrix regime (i.e. when both the number of samples and the dimension tends to infinity at the same rate) has been established in \cite{zhang2014marchenko, couillet2015therandom}. 
Recently, \cite{draskovic2016new_tyler, draskovic2018new_insights} showed that their distribution can be very accurately approximated by a Wishart one of an equivalent Gaussian core model referred to as Gaussian Core Wishart Equivalent (GCWE).

Additionally, the eigenvalue decomposition (EVD) of $M$-estimators is required in numerous processes. Indeed, the eigenvectors of the scatter matrix are involved in probabilistic PCA algorithms \cite{croux00principalcomponent, zhao2006probabilistic}, as well as in the derivation of robust counterparts of low rank filters or detectors \cite{Rangaswamy2005detector, Ginolhac16}.
The eigenvalues of the scatter are used in model order selection \cite{stoica2004model, terreaux18robust}, functions of eigenvalues are involved in various applications such as regularization parameter selection \cite{ollila2014regularized, Kammoun18}, detection \cite{ciuonzo2017covarianceequality}, and classification \cite{bouveyron2014model}.
Hence accurately characterizing the distribution of the $M$-estimators EVD represents an interest, both from the points of view of performance analysis and optimal process design.
In this paper, we derive new asymptotic characterizations for the EVD parameters of scatter $M$-estimators in the general context of CES-distributed samples. For the eigenvalues, the eigenvectors and the principal subspace (i.e. the subspace spanned by the $r$ strongest eigenvectors), we derive:
\begin{itemize}
	\item The standard Gaussian asymptotic distribution. This result is obtained by extending the analysis of \cite{kollo1993asymptotics} (for the SCM) 
	and perturbation analysis of \cite{krim92operator, krim96projections} to the complex $M$-estimators. 
	This asymptotic analysis provides 
	an extension of \cite{tyler1981asymptotics_eigenvectors, croux00principalcomponent} since it gives the information about the covariance between the eigenvalues of an $M$-estimator and provides the exact structure of the asymptotic covariance and pseudo-covariance matrix of principal subspace. Also, contrary to \cite{tyler1981asymptotics_eigenvectors, croux00principalcomponent}, all the results in this paper are derived for complex data. 
	\item The convergence rate towards the EVD parameters of a GCWE by extending the results of  \cite{draskovic2016new_tyler, draskovic2018new_insights}. This result represents the main contribution in the sense that it quantifies when it is acceptable to directly plug-in well established results on the EVD of Wishart-distributed matrices for characterizing the EVD of $M$-estimators \cite{muirhead1982aspects, Zanella09}.
\end{itemize}
In the last part, we eventually give some examples where the proposed results can be leveraged. 
Concerning the eigenvectors and principal subspaces, we derive the performance of Low rank filters \cite{Ginolhac10} build from $M$-estimators. 
Regarding the eigenvalues, we address the complex issue of characterizing the intrinsic bias \cite{smith2005covariance} of $M$-estimators in CES distribution. This quantity has been studied in \cite{smith2005covariance} for the SCM in a Gaussian context thanks to the distribution of the eigenvalues of a Wishart matrix \cite{muirhead1982aspects}. Extending this analysis to $M$-estimators represents, at first sight, an intractable problem. However, our proposed GCWE allows to derive an accurate approximation of this quantity.

The rest of this paper is organized as follows. Section \ref{sec:estim} introduces the CES distributions and $M$-estimators. Section \ref{sec:eigen} contains the main results about eigenvalue decomposition of $M$-estimators. In Section \ref{sec:LR}, we introduce LR standard models and present main results about principal subspaces of $M$-estimators. In Section \ref{sec:simul} Monte Carlo simulations are presented in order to validate the theoretical results. In addition, examples of applications of the results are presented. Finally, some conclusions and perspectives are drawn in Section \ref{sec:conclu}.\\

\textit{Notations} - Vectors (resp. matrices) are denoted by bold-faced lowercase letters (resp. uppercase letters). $^T$, $^*$, $^H$ and $^+$ respectively represent the transpose, conjugate, Hermitian operator and pseudo-inverse of a matrix. i.i.d. stands for ``independent and identically distributed'', w.r.t. for ``with respect to'' and $\sim$ means ``is distributed as''. $\overset{d}{=}$ stands for ``shares the same distribution as'', $\overset{d}{\to}$ denotes convergence in distribution and $\otimes$ denotes the Kronecker product. $\vec(\cdot)$ is the operator which transforms a matrix $ p \times n$ into a vector of length $pn$, concatenating its $n$ columns into a single column. Moreover, $\Iden_p$ is the $p \times p$ identity matrix, $\0b$ the matrix of zeros with appropriate dimension and $\Kb$ is the commutation matrix (square matrix with appropriate dimensions) which transforms $\vec(\Ab)$ into $\vec(\Ab^T)$, i.e. $\Kb\,\vec(\Ab) = \vec(\Ab^T)$. 
$\mathcal{H}_M^{++}$ is the set of Hermitian positive definite matrices.
The set of semi-unitary matrices is denoted as $\mathcal{U}_r^p = \left\{ \mathbf{U} \in \mathbb{C}^{p\times r} , \mathbf{U}^H \mathbf{U}  = \mathbf{I}_r\right\}$. Finally, $\mathcal{GCN}\pg\0b, \Vb, \Wb \pd$ denotes the zero-mean non-circular improper complex normal distribution with covariance matrix $\Vb$ and pseudo-covariance matrix $\Wb$ \cite{ollila2012complex}.

\section{Background}
\label{sec:estim}

\subsection{CES distributions}
Complex Elliptically Symmetric (CES) distributions form a general family of circular multivariate distributions \cite{ollila2012complex}. The probability density functions (PDFs) of a CES distribution can be written as 
\begin{eqnarray}
\label{pdfel}
f_{\zb}(\zb)&=&C |\Sig |^{-1}\,g_{\zb}\pg(\zb-\mub)^H \Sig^{-1}(\zb-\mub)\pd 
\end{eqnarray}
where $C$ is a normalisation constant and $g_{\zb}:[0,\infty)\rightarrow [0,\infty)$ is any function (called the density generator) ensuring Eq. \eqref{pdfel}  defines a PDF. These CES distributions will be denoted by $\mathcal{CES}\pg\mub,\Sig,g_{\zb}\pd$. The Complex Normal (Gaussian) distribution is a particular case of CES distributions in which $g_{\zb}(z)=e^{-z}$ and $C=\pi^{-p}$. We denote this distribution as $\zb \sim \mathcal{CN}(\mub,\Sig)$ (see \cite{ollila2012complex} and Section \ref{sec:simul} for more examples of CES distributions).

\textit{Gaussian-cores representation}
In \cite{draskovic2018new_insights} the so-called Gaussian-core model has been used as an alternative to the classic stochastic representation \cite{Yao73} of CES-distributed vectors. A random vector $\zb \sim \mathcal{CES}(\0b,\Sig ,g_{\zb})$ can be represented as
\begin{equation}
\label{gauscore}
\zb\overset{d}{=}\frac{\sqrt{\mathcal{Q}}}{\|\gb\|}\Ab\gb
\end{equation}
where $\Sig =\Ab\Ab^H$ is a factorization of $\Sig$ and $\gb \sim \mathcal{CN}(\0b,\Iden)$. $\mathcal{Q}$ is a non-negative real random variable, called the modular variate, independent of $\gb$ with a PDF depending only on $g_{\zb}$. We refer to $\xb=\Ab\gb$ as the Gaussian-core of $\zb$.

\subsection{$M$-estimators and SCM}
Let $(\zb_1, \hdots, \zb_n)$ be an $n$-sample of $p$-dimensional complex i.i.d. vectors with $\zb_i \sim \mathcal{CES}(\0b,\Sig, g_{\zb})$. An $M$-estimator, denoted by $\widehat \Sig$, is defined by the solution of the following $M$-estimating equation 
\begin{equation}
\label{mest}
\widehat \Sig = \frac{1}{n}\sum_{i=1}^{n}u(\zb_i^H \widehat \Sig^{-1}\zb_i)\zb_i\zb_i^H
\end{equation}
where $u$ is any real-valued weight function on $[0,\infty)$ that respects Maronna's conditions (ensuring existence and uniqueness)\cite{Maronna76}. The theoretical scatter matrix $M$-functional is defined as a solution of
\begin{equation} 
\label{mfunc}
\mathbb{E}[u(\zb^H \Sig_{\sigma}^{-1}\zb)\zb\zb^H] =\Sig_{\sigma}.
\end{equation}
The $M$-functional is proportional to the true scatter matrix parameter $\Sig$ as $\Sig_{\sigma}=\sigma^{-1}\Sig$, where the scalar factor $\sigma>0$ can be found by solving
\begin{equation}
\label{sig}
\mathbb{E}[\Psi(\sigma t)]=p
\end{equation}
with $\Psi(\sigma t)=u(\sigma t)\sigma t$ and $t=\zb^H \widehat \Sig^{-1}\zb$.

The sample covariance matrix (SCM) \cite{statistics1999} is given by
\begin{equation}
\label{scm}
\widehat{\Sig}_{\rm SCM} = \frac{1}{n}\sum_{i=1}^{n}\zb_i\zb_i^H.
\end{equation}
The SCM can be considered as a ``limited case'' of Eq. \eqref{mest} when $u(\zb_i^H \widehat \Sig^{-1}\zb_i)=1$. This estimator is usually used when the data is assumed to be Gaussian-distributed, since it is the Maximum Likelihood Estimator (MLE) in that case. Note that for the SCM, \eqref{mest} becomes explicit which makes this estimator very convenient for statistical analysis. Indeed, for $\zb\sim\mathcal{CN}\pg\0b, \Sig\pd$, it follows a Wishart distribution with well-known properties \cite{muirhead1982aspects}. However, since the SCM is not robust, it can perform very poorly in comparison to $M$-estimators in CES framework.

\subsection{Standard Asymptotic Regime}
Let $(\zb_1,\hdots,\zb_n)$ be an $n$-sample of $p$-dimensional complex independent vectors with $\zb_i\sim\pg\0b,\Sig,g_{\zb}\pd$, $i=1,\hdots,n$. We consider the complex $M$-estimator $\widehat \Sig$ that
verifies Eq. \eqref{mest} and follows Maronna's conditions \cite{Maronna76}, and we denote $\Sig_{\sigma}$ the solution of Eq. \eqref{mfunc}.

\begin{Thm}
	\label{thm-m}
	The asymptotic distribution of $\widehat{\Sig}$ is given by \cite{mahot2013asymptotic, ollila2012complex} as
\begin{equation*}
\sqrt{n}\vec\pg\widehat \Sig- \Sig_{\sigma}\pd \overset{d}{\to} \mathcal{GCN}\pg\0b,\Cb,\Pb\pd
\end{equation*}
	where the asymptotic covariance and pseudo-covariance matrices are
\begin{equation}
\label{asymp-MC}
\left\{
\begin{array}{l}
\Cb=\vartheta_1 \Sig_{\sigma}^{T}\otimes \Sig_{\sigma}+\vartheta_2\vec\pg \Sig_{\sigma}\pd\vec\pg\Sig_{\sigma}\pd^{H} , \\
\Pb=\vartheta_1\pg\Sig_{\sigma}^{T}\otimes \Sig_{\sigma}\pd\Kb+\vartheta_2\vec\pg \Sig_{\sigma}\pd\vec\pg\Sig_{\sigma}\pd^{T}.
\end{array}
\right.
\end{equation}
	The constants $\vartheta_1 > 0$ and $\vartheta_2>-\vartheta_1/p$ are given by
	\begin{equation}
	\label{coefSA}
	\begin{array}{l}
	\vartheta_1=c_M^{-2}a_Mp(p+1)  \\
	\vartheta_2=(c_M-p^2)^{-2}(a_M-p^2)-c_M^{-2}a_M(p+1)
	\end{array}
	\end{equation}
	where
	\begin{eqnarray*}
		a_M&=&E[\Psi^2(\sigma \mathcal{Q})] \label{a_M} \\
		c_M&=&E[\Psi'(\sigma \mathcal{Q})\sigma \mathcal{Q}]+p^2.
	\end{eqnarray*}
\end{Thm}
with $\mathcal{Q}$ defined in Eq. \eqref{gauscore}.
\begin{remark}
	Note that for the SCM built with Gaussian-ditributed data $\vartheta_1=1$ and $\vartheta_2=0$ with $\Sig_{\sigma}=\Sig$.
\end{remark}

\subsection{Gaussian-Core Wishart Equivalent (GCWE)}
\label{subsec:GCWE}
The asymptotic distribution of the difference between an $M$-estimator and the SCM built with Gaussian-cores of CES data (Eq. \eqref{gauscore}), has been recently derived in \cite{draskovic2018new_insights}.
 
\textit{Assumed Gaussian-core model:} Let us assume $n$ measurements $(\zb_1,\hdots,\zb_n)$  where  $\zb_i=\sqrt{\mathcal{Q}_i}/\|\gb_i\|\Ab\gb_i$  following $\mathcal{CES}\pg\0b,\Sig,g_{\zb}\pd$, $i=1,\hdots,n$, where
\begin{itemize}
	\item $\widehat\Sig$ is an $M$-estimator built with $(\zb_1,\hdots,\zb_n)$ using Eq. \eqref{mest},
	\item $\widehat\Sig_{\rm GCWE}=\frac{1}{n}\sum_{i=1}^{n}\xb_i\xb_i^H$ is the SCM built with Gaussian-cores  $\xb_i=\Ab\gb_i \sim \mathcal{CN}(\0b,\Sig)$ of $\zb_i$, $i=1,\hdots,n$, given by Eq. \eqref{gauscore} that represent only fictive data used for theoretical purposes.
\end{itemize}
Hereafter, we always consider the same model.

\begin{Thm}
	\label{thmGCWE}
	Let $\sigma$ be the solution of Eq. \eqref{sig}. Then, the asymptotic distribution of $\sigma \widehat \Sig - \widehat \Sig_{GCWE}$ is given by \cite{draskovic2018new_insights} 
	\begin{equation}
	\label{complex_thm}
	\sqrt{n}\vec\pg\sigma\widehat \Sig-\widehat \Sig_{GCWE}\pd \overset{d}{\to}\mathcal{GCN} \pg\0b,\widetilde{\Cb},\widetilde{\Pb}\pd
	\end{equation}
	where $\widetilde{\Cb}$ and $\widetilde{\Pb}$ are defined by
	\begin{eqnarray}
	\widetilde{\Cb} &=&\sigma_1 \Sig^{T}\otimes \Sig+\sigma_2\vec( \Sig)\vec( \Sig)^{H} ,  \notag  \\
	\widetilde{\Pb}&=&\sigma_1 \pg\Sig^{T}\otimes \Sig\pd\Kb+\sigma_2 \vec( \Sig)\vec( \Sig)^{T}
	\end{eqnarray}
	with $\sigma_1$ and $\sigma_2$ given by
	\begin{eqnarray}
	\label{coefGCWE}
	\sigma_1&=&\frac{ap(p+1)+c(c-2b)}{c^2} \notag \\
	\sigma_2&=&\frac{a-p^2}{(c-p^2)^2}-\frac{a(p+1)}{c^2}+2\frac{p(c-b)}{c(c-p^2)}
	\end{eqnarray}
	where
	\begin{eqnarray*}
		a&=&E[\Psi^2(\sigma \mathcal{Q})] \notag \\
		b&=&E[\Psi(\sigma \mathcal{Q})\|\gb\|^2]  \notag \\
		c&=&E[\Psi'(\sigma \mathcal{Q})\sigma \mathcal{Q}]+p^2.
	\end{eqnarray*}

%	\begin{eqnarray}
%	\label{sigma-complex}
%	\sigma_1&=&\frac{am(m+1)+c(c-2b)}{c^2} \notag \\
%	\sigma_2&=&\frac{a-m^2}{(c-m^2)^2}-\frac{a(m+1)}{c^2}+2\frac{m(c-b)}{c(c-m^2)}
%	\end{eqnarray}
%	where $a=E[\omega^2(\sigma t_1)]$, $b=E[\omega(\sigma t_1)t_2]$ and $c=E[\omega'(\sigma t_1)\sigma t_1]+m^2$.
\end{Thm}

An important note is that these factors are much smaller than the ones in the regular asymptotic regime (Eq. \eqref{coefSA}) meaning that the behavior of an $M$-estimator can be accurately approximated with the behavior of the corresponding Wishart-distributed matrix.

\section{Asymptotics of $M$-estimators' eigenvalue decomposition}
\label{sec:eigen}

The EigenValue Decomposition (EVD) of a (scatter) matrix $\Sig$ is defined as
\begin{equation}
\mathbf{\Sigma} \overset{\rm{EVD}}{=} \mathbf{U} \mathbf{\Lambda} \mathbf{U}^H
\end{equation}
with
\begin{equation}
\mathbf{U}^H\mathbf{U}=\Iden
\end{equation}
where $\mathbf{U} = \left[ \mathbf{u}_1 , \ldots, \mathbf{u}_p \right] \in \mathcal{U}_p^p$ 
and 
$\mathbf{\Lambda} = {\rm diag}( \boldsymbol{\lambda} )$,
$\boldsymbol{\lambda}  = \left[ \lambda_1 , \ldots, \lambda_p  \right] \in \mathbb{R}^p$.
In the following we assume ordered eigenvalues $\lambda_1>\hdots>\lambda_p >0$ .

We define the operators $\thb_j$ and $\phi_j$ returning respectively the $j^{\rm th}$ eigenvector and eigenvalue as
\begin{equation}
\left\{
\begin{aligned}
\ub_j&=&\thb_j\pg\Sig\pd, \\
\lambda_j&=&\phi_j\pg\Sig\pd. 
\end{aligned}
\right.
\end{equation}

Let us again assume the Gaussian-core model proposed in Section \ref{subsec:GCWE} and
\begin{equation}
\left\{
\begin{array}{ll}
\widehat{\ub}_j^M=\thb_j\pg\widehat\Sig\pd & \widehat{\ub}_j^{\rm GCWE}=\thb_j\pg\widehat\Sig_{\rm GCWE}\pd,\\
\widehat{\lambda}_j^M=\phi_j\pg\widehat\Sig\pd & \widehat{\lambda}_j^{\rm GCWE}=\phi_j\pg\widehat\Sig_{\rm GCWE}\pd.
\end{array}
\right.
\end{equation}
with $\phb=\cg\phi_1,\hdots,\phi_p\cd$.
In the following we derive the asymptotic distribution for these quantities.

\begin{Thm}[Standard asymptotic regime]
\label{thm-1}
Let $\hat{\mathbf{\Sigma}}$ be a scatter $M$-estimator with
$\hat{\mathbf{\Sigma}}   \overset{\rm{EVD}}{=} \hat{\mathbf{U}} {\rm diag} ( \hat{\boldsymbol{\lambda}} ) \hat{\mathbf{U}}^H $.
The asymptotic distribution of the eigenvalues and eigenvectors is characterized by
\begin{equation}
\label{eig-MC}
\left\{
\begin{array}{l}
\sqrt{n}  \pg \sigma\widehat{\lamb}^{M}-\lamb\pd\overset{d}{\to} \mathcal{N}\pg\0b, \vartheta_1\Lamb^2+\vartheta_2\lamb\lamb^T\pd, \\
\sqrt{n} \Pib_j^{\bot} \widehat{\ub}_{j}^{M}  \overset{d}{\to} \mathcal{CN}\pg\0b,\Xib_{j}\pd.
\end{array}
\right.
\end{equation}
where 
\begin{equation}
\label{ximat}
	\Xib_{j}=\vartheta_1\lambda_j \Ub\Lamb(\lambda_j\Iden-\Lamb)^{+^2}\Ub^H
\end{equation} 
with $\Pib_j^{\bot}=\Iden -\ub_j\ub_j^H$ and $\vartheta_1$, $\vartheta_2$ given by Eq. \eqref{coefSA}.
\end{Thm}
\begin{IEEEproof}
	See Appendix \ref{app1}.
\end{IEEEproof}

\begin{Thm}[GCWE]
\label{thm-2}
Asymptotic distribution of the difference between the eigenvalues and eigenvectors of an $M$-estimators and GCWE is given by
\begin{equation}
\label{eigdiff-MC}
\left\{
\begin{array}{l}
\sqrt{n} \pg\sigma \widehat{\lamb}^{M}-\widehat{\lamb}^{\rm GCWE}\pd\overset{d}{\to} \mathcal{N}\pg\0b, \sigma_1\Lamb^2+\sigma_2\lamb\lamb^T\pd, \\
\sqrt{n}\Pib_j^{\bot}\pg \widehat{\ub}_{j}^{M}-\widehat{\ub}_j^{\rm GCWE}\pd\overset{d}{\to} \mathcal{CN}\pg\0b, \sigma_1/\vartheta_1\Xib_{j}\pd.
\end{array}
\right.
\end{equation}
with $\Xib_{j}$ and $\sigma_1,\sigma_2$ given by Eqs. \eqref{ximat} and \eqref{coefGCWE}, respectively.
\end{Thm}
\begin{IEEEproof}
	See Appendix \ref{app2}.
\end{IEEEproof}

\begin{remark}
\begin{itemize}
\item The results given in Theorem \ref{thm-1} are interesting since, besides the variance of each eigenvalue, they provide the correlation between them. Note that for a Wishart-distributed matrix this correlation is equal to zero, as shown in \cite{kollo1993asymptotics} for real case.
Conversely, we showed that the eigenvalues of an $M$-estimator are asymptotically correlated, as stated in \cite{croux00principalcomponent} (but not explicitly characterized).
This correlation depends on the second scale parameter $\vartheta_2$. 
Concerning the eigenvectors, note that the covariance depends only on $\vartheta_1$ since  $\thb_j$ is scale invariant w.r.t. to the covariance matrix (see \cite{mahot2013asymptotic} for more details).\\

\item Theorem \ref{thm-2} characterizes the asymptotic variance of the EVD of an $M$-estimator compared to the one of its GCWE.
It shows that their covariance structure is the same as for the standard asymptotic regime. However the scale $\sigma_1$ is much smaller than $\vartheta_1$, especially when $p$ increases. 
Therefore, the GCWE provides a better asymptotic characterization of the $M$-estimator's EVD.
\end{itemize}
\end{remark}

\section{Asymptotics of $M$-estimators' principal subspace}
\label{sec:LR}

Consider the special case of low-rank plus identity (also referred to as factor model)
\begin{equation}
\label{lrmodel}
\mathbf{\Sigma}  = \mathbf{\Sigma}_r + \gamma^2 \mathbf{I}_p
\end{equation}
with
\begin{equation}
\label{lrsig}
\mathbf{\Sigma}_r  = \Ub_r \mathbf{\Lambda}_r \Ub_r^H
\end{equation}
with $\Ub_r \in \mathcal{U}_r^p$ and $\mathbf{\Lambda} \in \mathbb{C}^{r \times r}$.
The principal subspace is defined as
\begin{equation}
\label{proj}
\mathbf{\Pi}_r = \Ub_r \Ub_r^H.
\end{equation}

Let us consider an $M$-estimator built with $(\zb_1,\hdots,\zb_n)$ where $\zb_i\sim\mathcal{CES}\pg\0b,\mathbf{\Sigma}_r + \gamma^2 \mathbf{I}_p\pd$ and the GCWE built with fictive data $(\xb_1,\hdots,\xb_n)$ given by Eq. \eqref{gauscore} where $\xb_i\sim\mathcal{CN}\pg\0b,\mathbf{\Sigma}_r + \gamma^2 \mathbf{I}_p\pd$, $i=1,\hdots, n$. Assume then that $\widehat\Ub_r^M$ is the estimate of $\Ub_r$ obtained with the $M$-estimator, i.e., $\widehat\Ub_r^M=\cg\thb_1\pg\widehat{\Sig}\pd,\hdots,\thb_r\pg\widehat{\Sig}\pd\cd$, while $\widehat\Ub_r^{\rm GCWE}$ is its estimate obtained with the GCWE, i.e., $\widehat\Ub_r^{\rm{GCWE}}=\cg\thb_1\pg\widehat{\Sig}_{\rm{GCWE}}\pd,\hdots,\thb_r\pg\widehat{\Sig}_{\rm{GCWE}}\pd\cd$. Therefore, one can construct the following projectors 
\begin{equation}
\label{proj-M}
\left\{
\begin{array}{l}
\widehat{\Pib}_r^M=\widehat\Ub_r^M\pg\widehat\Ub_r^M\pd^H ,\\
\widehat{\Pib}_r^{\rm{GCWE}}=\widehat\Ub_r^{\rm{GCWE}}\pg\widehat\Ub_r^{\rm{GCWE}}\pd^H.
\end{array}
\right.
\end{equation}

\begin{Thm}[Standard asymptotic regime]
\label{thm-3}
Let $\widehat\Pib_r^M$ the estimate of the projector $\Pib_r$ obtained using an $M$-estimator defined in Eq. \eqref{proj-M}. The asymptotic distribution of $\widehat\Pib_r^M$ is given by
\begin{equation}
\label{proj-MC}
\sqrt{n}\vec\pg\widehat\Pib_r^M-\Pib_r\pd\overset{d}{\to} \mathcal{GCN}\pg\0b,\vartheta_1\Sig_{\Pib}, \vartheta_1\Sig_{\Pib}\Kb\pd, 
\end{equation}
where
\begin{equation}
\label{sigpi}
	\Sig_{\Pib}=\Ab^T\otimes\Bb+\Bb^T\otimes\Ab
\end{equation}
with $\Ab=\Ub_r\pg\gamma^2\Lamb_r^{-2}+\Lamb_r^{-1}\pd\Ub_r^H$, $\Bb=\gamma^2\Pib_r^{\bot}$ and $\vartheta_1,\vartheta_2$ given by Eq. \eqref{coefSA}.
\end{Thm}
\begin{IEEEproof}
	See Appendix \ref{app3}.
\end{IEEEproof}

\begin{Thm}[GCWE]
\label{thm-4}
Let $\widehat\Pib_r^M$ and $\widehat\Pib_r^{\rm{GCWE}}$ be the estimates of the projector $\Pib_r$ defined in Eq. \eqref{proj-M}. Then, the asymptotic distribution of $\widehat\Pib_r^M$ is given by
\begin{equation}
\label{projdiff-MC}
\sqrt{n}\vec\pg\widehat\Pib_r^M-\widehat\Pib_r^{\rm{GCWE}}\pd\overset{d}{\to} \mathcal{GCN}\pg\0b,\sigma_1\Sig_{\Pib}, \sigma_1\Sig_{\Pib}\Kb\pd
\end{equation}
with $\Sig_{\Pib}$ and $\sigma_1, \sigma_2$
given by Eqs. \eqref{sigpi} and \eqref{coefGCWE}, respectively.
\end{Thm}
\begin{IEEEproof}
	See Appendix \ref{app4}.
\end{IEEEproof}

\begin{remark} 
Theorem \ref{thm-3} ($\resp$ \ref{thm-4}) extend the results of Theorem \ref{thm-1} ($\resp$ \ref{thm-2}) to the principal subspace of $M$-estimators, which is a parameter of significant interest.  We can draw the same conclusions as previously, notably, that the GCWE provides a better asymptotic characterization of this parameter.
\end{remark}

\section{Simulations and examples}
\label{sec:simul}

\subsection{Parameters setup}

In order to validate the theoretical results we draw zero-mean $t$-distributed data with $d$ degrees of freedom (DoF) whose PDF is given by Eq. \eqref{pdfel} with
\begin{equation}
	g_{\zb}(x)=\pg 1 + \frac{2x}{d}\pd^{-\pg p+d/2\pd}
\end{equation}
and $C_t=2^p\Gamma(p+\frac{d}{2})/[(\pi d)^p \Gamma(\frac{d}{2})]$. The corresponding stochastic representation is given by Eq. \eqref{gauscore} for $\mathcal{Q} \sim pF_{2p,d}$.

The DoF parameter is set to 3. The dimension of the data is $p=20$. The scatter matrix is Toeplitz, i.e. elements are defined by $\Sigma_{jk}=\rho^{|j-k|}$, $j,k = 1, \hdots,p$, with correlation coefficient $\rho$ set to $0.9(1+\sqrt{-1})/\sqrt{2}$. 

In order to carry out the simulations we will use the Student's $M$-estimator that is the MLE for Student's $t$-distribution and can be obtained as solution of Eq. \eqref{mest} for
\begin{equation}
	u\pg x \pd= \frac{2p+d}{d+2x}.
\end{equation}

In this context, the parameters for asymptotic distribution of Student's $M$-estimator are given in Table \ref{stcoeff}.
%\begin{equation}
%\label{student}
%\left\{
%\begin{array}{ll}
%\vartheta_1=\frac{m+\frac{d}{2}+1}{m+\frac{d}{2}} & \vartheta_2=\frac{2}{d}\frac{m+\frac{d}{2}+1}{m+\frac{d}{2}}, \\
%\sigma_1=\frac{1}{m+\frac{d}{2}} & \sigma_2=\frac{2}{d}\frac{m+\frac{d}{2}+1}{m+\frac{d}{2}}.
%\end{array}
%\right.
%\end{equation}

	\begin{table}[!h]
		\begin{center}
			\begin{tabular}{ |c|c| } 
			\hline
			~&\\
			Standard regime & Gaussian equivalent  \\ &\\\hline
			&\\
			$\vartheta_1=\cfrac{p+d/2+1}{p+d/2}$ & $\sigma_1=\cfrac{1}{p+d/2}$  \\ &\\ \hline
				&\\
			$\vartheta_2=\cfrac{2}{d}\times\cfrac{p+d/2+1}{p+d/2}$ & $\sigma_2=\cfrac{2}{d}\times\cfrac{p+d/2+1}{p+d/2}$  \\ 
			&\\ \hline
		\end{tabular}
	\caption{\label{stcoeff} Coefficients $\vartheta_1$, $\vartheta_2$, $\sigma_1$ and $\sigma_2$ for Student's $M$-estimator}
\end{center}
	\end{table}

\subsection{Experiments for eigenvalues}
\subsubsection{Validation of theoretical results}
Let us consider $\widehat{\lamb}^t=\phb\pg\widehat{\Sig}_t\pd$ where $\widehat{\Sig}_t$ is the Student's $M$-estimator and $\widehat{\lamb}^{\rm{GCWE}}=\phb\pg\widehat{\Sig}_{\rm{GCWE}}\pd$, where $\widehat{\Sig}_{\rm{GCWE}}$ is the SCM built with the Gaussian kernels of the observed data (GCWE).

\begin{figure}[!h]
\begin{center}
\begin{tikzpicture}[scale=0.8]

\begin{axis}[%
%width=0.951\fwidth,
%height=0.75\fwidth,
%at={(0\fwidth,0\fwidth)},
scale only axis,
xmode=log,
xmin=40,
xmax=2000,
xminorticks=true,
xlabel style={font=\color{white!15!black}},
xlabel={$n$},
ymin=-15,
ymax=10,
ylabel style={font=\color{white!15!black}},
ylabel={Mean Squared Error (dB)},
axis background/.style={fill=white},
title style={font=\bfseries},
title={},
xmajorgrids,
xminorgrids,
ymajorgrids,
legend style={legend cell align=left, align=left, draw=white!15!black}
]
\addplot [color=red, mark=star, mark options={solid, red}]
  table[row sep=crcr]{%
40	9.6252394431301\\
62	7.44024583261122\\
95	4.71172761701018\\
147	2.73889390220967\\
228	0.670871859894525\\
352	-1.18320935421492\\
543	-3.32127836091408\\
838	-5.16551542350919\\
1295	-7.05896861376744\\
2000	-8.91195623818191\\
};
\addlegendentry{MSE$\pg\widehat{\lamb}^t-\lamb\pd$}

\addplot [color=black, dashed]
  table[row sep=crcr]{%
40	8.04819491734646\\
62	6.14487793564355\\
95	4.29155877773761\\
147	2.39562148314432\\
228	0.489446360621548\\
352	-1.39663180415522\\
543	-3.27920346526238\\
838	-5.16364535567668\\
1295	-7.05390285354662\\
2000	-8.94150512601373\\
};
\addlegendentry{$\mathcal{T}\pg\widehat{\lamb}^t-\lamb\pd$}

\addplot [color=blue, mark=star]
  table[row sep=crcr]{%
40	6.51676517993395\\
62	3.97954798239548\\
95	1.50925412281768\\
147	-1.01868997753239\\
228	-2.92395692511326\\
352	-4.82383769503098\\
543	-6.87049476855998\\
838	-8.86671963512157\\
1295	-10.8264154808781\\
2000	-12.6141515781282\\
};
\addlegendentry{MSE$\pg\widehat{\lamb}^t-\widehat{\lamb}^{GCWE}\pd$}

\addplot [color=black]
  table[row sep=crcr]{%
40	4.34908206662852\\
62	2.44576508492561\\
95	0.592445927019667\\
147	-1.30349136757362\\
228	-3.20966649009639\\
352	-5.09574465487317\\
543	-6.97831631598033\\
838	-8.86275820639462\\
1295	-10.7530157042646\\
2000	-12.6406179767317\\
};
\addlegendentry{$\mathcal{T}\pg\widehat{\lamb}^t-\widehat{\lamb}^{GCWE}\pd$}

\end{axis}
\end{tikzpicture}%
\end{center}
\caption{Empirical (MSE) and theoretical ($\mathcal{T}$) mean squared error on eigenvalues: Results for the standard asymptotic regime $\pg\widehat{\lamb}^t-\lamb\pd$ and Gaussian equivalent $\pg\widehat{\lamb}^t-\widehat{\lamb}^{\rm{GCWE}}\pd$; $t$-distributed data with $p = 20$, $d = 3$.
}
\label{fig1}
\end{figure}
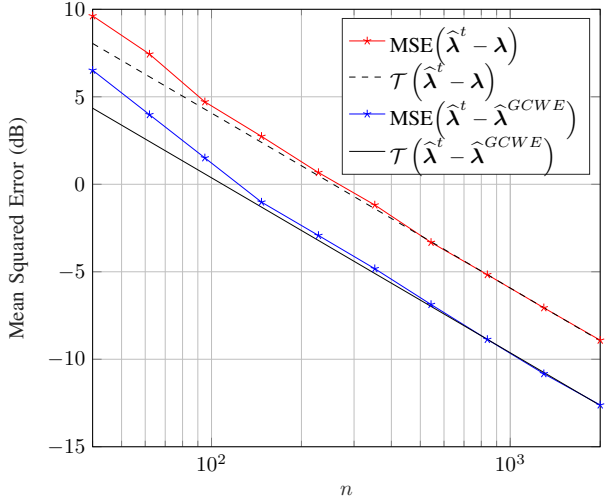

Figure \ref{fig1} displays the empirical mean squared error (MSE) of $\widehat{\lamb}^t$ when comparing to $\lamb$ and $\widehat{\lamb}^{\rm{GCWE}}$, denoted as MSE$\pg\widehat{\lamb}^t-\lamb\pd$ and MSE$\pg\widehat{\lamb}^t-\widehat{\lamb}^{\rm{GCWE}}\pd$ respectively. These quantities are compared to the corresponding asymptotic theoretical values, i.e. the traces of the asymptotic covariance matrices given in Eqs. \eqref{eig-MC} and \eqref{eigdiff-MC}, denoted as $\mathcal{T}\pg\widehat{\lamb}^t-\lamb\pd=\rm Tr\pg\vartheta_1\Lamb^2+\vartheta_2\lamb\lamb^T\pd$ and $\mathcal{T}\pg\widehat{\lamb}^t-\widehat{\lamb}^{\rm{GCWE}}\pd=\rm Tr\pg\sigma_1\Lamb^2+\sigma_2\lamb\lamb^T\pd$. The horizontal scale represent the number of observations $n$.

First, we observe from Figure \ref{fig1} that the empirical results (red and blue curves) tend to the corresponding theoretical ones (black curves) as $n$ increases. Another remark is that the error of $\widehat{\lamb}^t$ is much smaller when comparing to $\widehat{\lamb}^{\rm{GCWE}}$ than to $\lamb$. This support the idea that the distribution of the eigenvalues of an $M$-estimator (in this case Student's MLE) is better approximated with the one of the eigenvalues of the GCWE, then with the asymptotic Normal distribution based on the eigenvalues of the true scatter matrix.

\subsubsection{Application}

In \cite{smith2005covariance} were derived an Intrinsic (i.e. Riemannian Manifold oriented) counterpart of the Cram\'er-Rao inequality.
In the context of covariance matrix estimation, these results allows notably to bound the expected natural Riemannian distance (rather than the Frobenius norm):
\begin{equation*}
{d}^2_{nat} 
\left( \mathbf{\Sigma}_1 , \mathbf{\Sigma}_2\right)
=
\left\| \ln \left(  
 \mathbf{\Sigma}_1^{-1/2} \mathbf{\Sigma}_2
 \mathbf{\Sigma}_1^{-1/2}
\right) \right\|_F^2
= 
\sum_{j=1}^p \ln^2  \lambda^d_j
\end{equation*}
where $\lambda^d_j$ is the $j^{\text{th}}$ eigenvalue of $\mathbf{\Sigma}_1^{-1} \mathbf{\Sigma}_2$.
This analysis also reveals unexpected and hidden properties of estimators, such as a bias of the SCM w.r.t. the natural metric on $\mathcal{H}_M^{++}$.
In this scope, the biased Intrinsic Cram\'er-Rao bound (CRLB) is established for the SCM in a Gaussian context as \cite[Corollary 5]{smith2005covariance}:
\begin{equation}
\label{eq:biased_clrb_scm}
\mathbb{E} 
\left[
{d}^2_{nat} 
\left( \hat{\mathbf{\Sigma}}_{\rm GCWE} , \mathbf{\Sigma} \right)
\right]\geq
\frac{p^2}{n}
+
p \eta \left( p, n  \right)^2 .
\end{equation}
The term $\eta \left( p, n  \right)$ in \eqref{eq:biased_clrb_scm} is related to the intrinsic bias (IB) of the SCM given in \cite[Theorem 7]{smith2005covariance} by 
\begin{equation} 
\mathbb{E} \left[ \exp^{-1}_\mathbf{\Sigma} \hat{\mathbf{\Sigma}}_{\rm GCWE} \right]
=
{- \eta(p,n)}  \mathbf{\Sigma}
\end{equation}
with
\begin{equation}
\label{eta}
\begin{aligned}
& {\eta(p,n)} & = ~&
\frac{1}{p}
\left\{ 
p \ln n
+ p
-
\psi(n-p+1)   \right.
 \\& & & ~~+
(n-p+1) \psi ( n -p +2)
 \\
& & &  ~~+ \left.
\psi( n+1 )
-
(n+1) \psi( n+2 )
\right\}
\end{aligned}
\end{equation}
and where $\psi(x) = \Gamma' (x) / \Gamma(x)$ is the digamma function.

For CES-distributed samples, the CRLB on $d_{nat}^2$ has been derived in \cite{breloy2018intrinsic} for any unbiased estimator $\hat{\mathbf{\Sigma}}$ as 
\begin{equation}
\mathbb{E} \left[ d_{nat}^2 \left( \hat{\mathbf{\Sigma}}, \mathbf{\Sigma}\right)\right] \geq \frac{p^2-1}{n\alpha} + (n(\alpha+ p \beta))^{-1}
.
\label{eq:intrinsic_crb_ces}
\end{equation}
with $\alpha =  \left( 1   -  \frac{\mathbb{E} \left[ \mathcal{Q}^2 \phi' \left( \mathcal{Q}  \right) \right]  
}{p(p+1)}   \right)
$ and $\beta =  {\alpha -1} $.
Extending the Corollary 5 of \cite{smith2005covariance} in this context would requires to derive the intrinsic bias of an $M$-estimator obtained with CES-distributed samples.
The problem appears intractable since this result is mainly obtained thanks to the distribution of the eigenvalues of a Wishart-distributed matrix. 
However the GCWE equivalent from Theorem \ref{thm-2} (as well as the previous simulation results) gives a reasonable theoretical ground for the approximation
$\mathbb{E}  [ \exp^{-1}_\mathbf{\Sigma} \hat{\mathbf{\Sigma}}_{M}  ]
\simeq
{- \eta(p,n)}  \mathbf{\Sigma} $ for any $M$-estimator consistent in scale under the CES framework.
Hence, we can propose to incorporate an equivalent bias term in \eqref{eq:intrinsic_crb_ces} to obtain an accurate approximation of the biased intrinsic CRLB for $M$-estimators build form CES-distributed samples (AB CRLB).

Figure \ref{fig2} illustrates this point.
\begin{figure}[!t]
	\begin{center}
\begin{tikzpicture}[scale=0.8]

\begin{axis}[%
%width=0.951\fwidth,
%height=0.75\fwidth,
%at={(0\fwidth,0\fwidth)},
scale only axis,
xmode=log,
xmin=40,
xmax=2000,
xminorticks=true,
xlabel style={font=\color{white!15!black}},
xlabel={ $n$},
ymin=-25,
ymax=0,
ylabel style={font=\color{white!15!black}},
ylabel={$\eta\text{ (dB)}$},
axis background/.style={fill=white},
title style={font=\bfseries},
title={},
xmajorgrids,
xminorgrids,
ymajorgrids,
legend style={legend cell align=left, align=left, draw=white!15!black}
]
\addplot [color=black]
  table[row sep=crcr]{%
40	-0.736002021819353\\
62	-4.38459785462046\\
95	-7.06561911459239\\
147	-9.49805354490065\\
228	-11.823959005642\\
352	-14.1036315278504\\
543	-16.3476934087354\\
838	-18.5751991398003\\
1295	-20.7857394129351\\
2000	-22.9957931882183\\
};
\addlegendentry{Eq.\eqref{eta}}

\addplot [color=red, dashed, mark=asterisk, mark options={solid, red}]
  table[row sep=crcr]{%
40	-0.723573211732889\\
62	-4.36222642186444\\
95	-7.07134989320909\\
147 -9.50586046328803\\
228	-11.8388874854454\\
352	-14.0905158246663\\
543	-16.3333620168431\\
838	-18.6341161295309\\
1295	-20.6877305349758\\
2000	-23.2043613622588\\
};
\addlegendentry{GCWE-IB}

\addplot [color=blue,dashed, mark=asterisk]
  table[row sep=crcr]{%
40	-1.47053923096041\\
62	-4.66961574660991\\
95	-7.33887679014756\\
147	-9.63477895570941\\
228	-12.1715369588326\\
352	-14.4836863627537\\
543	-16.583039960827\\
838	-18.7512157575151\\
1295	-21.3623363192133\\
2000	-23.4815014148484\\
};
\addlegendentry{Student-IB}

\end{axis}
\end{tikzpicture}%
	\end{center}
	\caption{ Empirical intrinsic bias for Student's $M$-estimator (Student-IB) and the Gaussian core SCM (GCWE-IB) compared to the theoretical result obtained for the GCWE (Eq. \eqref{eta})	}
	\label{fig2}
\end{figure}
Indeed, it can be seen that the empirical intrinsic bias obtained with Student's $M$-estimator computed with $t$-distributed data coincides with the intrinsic bias based on the SCM built with corresponding Gaussian-cores and the theoretical result (Eq. \eqref{eta}). This once again confirms previous results and supports the proposed approximation. In addition, on Figure \ref{fig3}, the results for CRLB on $d_{nat}^2$ have been plotted. Empirical mean of the natural Riemannian distance of $\widehat{\Sig}_t$ (denoted as $\epsilon^N\pg\widehat{\Sig}_t\pd$) is compared to the theoretical CRLB valid for any unbiased estimator (Eq. \eqref{eq:intrinsic_crb_ces}) and recommended approximation equal to the sum of the latter and bias term from Eq. \eqref{eq:biased_clrb_scm}. As expected, one can see that by introducing the bias term AB CRLB approaches to $\epsilon^N\pg\widehat{\Sig}_t\pd$ and gives more accurate theoretical results for CRLB.
\begin{figure}[!t]
	\begin{center}
\begin{tikzpicture}[scale=0.8]

\begin{axis}[%
%width=0.951\fwidth,
%height=0.75\fwidth,
%at={(0\fwidth,0\fwidth)},
scale only axis,
xmode=log,
xmin=40,
xmax=2000,
xminorticks=true,
xlabel style={font=\color{white!15!black}},
xlabel={$n$},
ymin=-10,
ymax=20,
ylabel style={font=\color{white!15!black}},
ylabel={Mean Squared Error (dB)},
axis background/.style={fill=white},
title style={font=\bfseries},
title={},
xmajorgrids,
xminorgrids,
ymajorgrids,
legend style={legend cell align=left, align=left, draw=white!15!black}
]
\addplot [color=black]
table[row sep=crcr]{%
	40	13.0550268691518\\
	62	10.8365393729882\\
	95	8.64293988086161\\
	147	6.45450748609531\\
	228	4.26324857328647\\
	352	2.06118054779268\\
	543	-0.137521288550007\\
	838	-2.33812429209939\\
	1295	-4.53265065261788\\
	2000	-6.73308014014882\\
};
\addlegendentry{Eq. \eqref{eq:intrinsic_crb_ces}}

\addplot [color=red, mark=star]
table[row sep=crcr]{%
	40	15.3728404921762\\
	62	11.6965986574844\\
	95	9.07887077118305\\
	147	6.6953078836901\\
	228	4.40153043247951\\
	352	2.14207503825224\\
	543	-0.0895889154492986\\
	838	-2.30953806675338\\
	1295	-4.51550815558961\\
	2000	-6.72278934242344\\
};
\addlegendentry{AB CRLB}

\addplot [color=blue, mark=asterisk]
table[row sep=crcr]{%
	40	18.8490008746111\\
	62	13.3462188722874\\
	95	9.97514914164615\\
	147	7.2353420184573\\
	228	4.71494282153387\\
	352	2.33654159219055\\
	543	0.0241106449172536\\
	838	-2.23697735147104\\
	1295	-4.46724529569678\\
	2000	-6.68875862877802\\
};
\addlegendentry{$\epsilon^N\pg\widehat{\Sig}_t\pd$}

\end{axis}
\end{tikzpicture}%
	\end{center}
	\caption{ Empirical mean of $d^2_{\text{nat}}\pg\widehat{\Sig}_t,\Sig\pd$ denoted as $\epsilon^N\pg\widehat{\Sig}_t\pd$ versus theoretical CRLB for an unbiased estimator in the CES framework (Eq. \eqref{eq:intrinsic_crb_ces}) and approximated biased instrinsic CRLB (AB CRLB)}
	\label{fig3}
\end{figure}

\subsection{Eigenvectors and Principal Subspace}

\subsubsection{Validation of theoretical results}

\begin{figure}[!t]
\begin{center}
\begin{tikzpicture}[scale=0.8]

\begin{axis}[%
%width=0.951\fwidth,
%height=0.75\fwidth,
%at={(0\fwidth,0\fwidth)},
scale only axis,
xmode=log,
xmin=40,
xmax=2000,
xminorticks=true,
xlabel style={font=\color{white!15!black}},
xlabel={$n$},
ymin=-50,
ymax=-10,
ylabel style={font=\color{white!15!black}},
ylabel={Mean Squared Error (dB)},
axis background/.style={fill=white},
title style={font=\bfseries},
title={},
xmajorgrids,
xminorgrids,
ymajorgrids,
legend style={legend cell align=left, align=left, draw=white!15!black}
]
\addplot [color=red, mark=asterisk]
  table[row sep=crcr]{%
40	-14.4532483496455\\
62	-16.3073791144194\\
95	-18.200346452982\\
147	-20.1018923750139\\
228	-22.0508604124771\\
352	-23.8497568620567\\
543	-25.8019044896595\\
838	-27.7486762558872\\
1295	-29.5434063026575\\
2000	-31.4738650013034\\
};
\addlegendentry{MSE$\pg\widehat{\ub}_1^t-{\ub}_1\pd$}

\addplot [color=black, dashed]
  table[row sep=crcr]{%
40	-14.491627675226\\
62	-16.3949446569289\\
95	-18.2482638148348\\
147	-20.1442011094281\\
228	-22.0503762319509\\
352	-23.9364543967277\\
543	-25.8190260578348\\
838	-27.7034679482491\\
1295	-29.5937254461191\\
2000	-31.4813277185862\\
};
\addlegendentry{$\mathcal{T}\pg\widehat{\ub}_1^t-{\ub}_1\pd$}

\addplot [color=blue, mark=star]
  table[row sep=crcr]{%
40	-24.4261299297243\\
62	-28.0754243637205\\
95	-30.6488179204209\\
147	-32.874175674599\\
228	-35.1453772359999\\
352	-37.1955545050332\\
543	-39.1118512341676\\
838	-41.1804555991081\\
1295	-43.0607300085691\\
2000	-44.9416539254209\\
};
\addlegendentry{MSE$\pg\widehat{\ub}_1^t-\widehat{\ub}_1^{GCWE}\pd$}

\addplot [color=black]
  table[row sep=crcr]{%
40	-28.0134528563396\\
62	-29.9167698380425\\
95	-31.7700889959485\\
147	-33.6660262905417\\
228	-35.5722014130645\\
352	-37.4582795778413\\
543	-39.3408512389484\\
838	-41.2252931293627\\
1295	-43.1155506272327\\
2000	-45.0031528996998\\
};
\addlegendentry{$\mathcal{T}\pg\widehat{\ub}_1^t-\widehat{\ub}_1^{GCWE}\pd$}

\end{axis}
\end{tikzpicture}%
\end{center}
\caption{Empirical and theoretical mean squared error on eigevectors: Results for the first eigenvector in the standard regime $\pg\widehat{\ub}_1^t-{\ub}_1\pd$ and for the Gaussian equivalent $\pg\widehat{\ub}_1^t-\widehat{\ub}_1^{\rm{GCWE}}\pd$; $t$-distributed data with $p = 20$, $d = 3$.
}
\label{fig4}
\end{figure}

%\begin{figure}[!t]
%	%\setlength\fwidth{0.38\textwidth}
%	\begin{center}
%		\input{vecpseudo.tex}
%	\end{center}
%	\caption{ Frobenius norm of the pseudo-covariance matrix of eigenvectors: Results for the first eigenvector $\pg\widehat{\ub}_1^t-{\ub}_1\pd$ and for the Gaussian equivalent $\pg\widehat{\ub}_1^t-\widehat{\ub}_1^{SCM}\pd$; $t$-distributed data with $M = 20$, $d = 3$.	}
%	\label{fig5}
%\end{figure}

Figure \ref{fig4} illustrates the results for the first eigenvector of $\widehat{\Sig}_t$, $\widehat{\ub}_1^t=\thb_1\pg\widehat{\Sig}_t\pd$. It is apparent from the plotted curves that empirical results, ones again, coincides well with  the theoretical ones. Moreover, the figure shows a significant difference between the results for the standard regime and GCWE. This can be explained by the fact that the covariance matrix of the eigenvectors depends only on the first scale factor, contrary to the one of eigenvalues.
\begin{figure}[!h]
	\begin{center}
\begin{tikzpicture}[scale=0.8]
\pgfplotsset{every axis legend/.append style={fill=white,cells={anchor=west},at={(0.01,0.33)},anchor=north west}}
\begin{axis}[%
%width=0.951\fwidth,
%height=0.75\fwidth,
%at={(0\fwidth,0\fwidth)},
scale only axis,
xmode=log,
xmin=40,
xmax=2000,
xminorticks=true,
xlabel style={font=\color{white!15!black}},
xlabel={$n$},
ymin=-40,
ymax=-5,
ylabel style={font=\color{white!15!black}},
ylabel={Mean Squared Error (dB)},
axis background/.style={fill=white},
title style={font=\bfseries},
title={},
xmajorgrids,
xminorgrids,
ymajorgrids,
legend style={legend cell align=left, align=left, draw=white!15!black}
]
\addplot [color=red, mark=asterisk]
  table[row sep=crcr]{%
40	-5.94155085848802\\
62	-8.15427015534933\\
95	-10.1430557035226\\
147	-12.1151963939328\\
228	-14.0765082167151\\
352	-15.987950166168\\
543	-17.9082250337168\\
838	-19.8082445520373\\
1295	-21.6903167106703\\
2000	-23.5842964583753\\
};
\addlegendentry{\small MSE$\pg\widehat{\Pib}^t-\Pib\pd$}

\addplot [color=black, dashed]
  table[row sep=crcr]{%
40	-6.61196004979674\\
62	-8.51527703149966\\
95	-10.3685961894056\\
147	-12.2645334839989\\
228	-14.1707086065217\\
352	-16.0567867712984\\
543	-17.9393584324056\\
838	-19.8238003228199\\
1295	-21.7140578206898\\
2000	-23.6016600931569\\
};
\addlegendentry{\small $\mathcal{T}\pg\widehat{\Pib}^t-\Pib\pd$}

\addplot [color=blue, mark=star]
  table[row sep=crcr]{%
40	-17.1248690465664\\
62	-20.288135768791\\
95	-22.7481097990115\\
147	-25.0608204351454\\
228	-27.1719084835979\\
352	-29.2212153000141\\
543	-31.194376550667\\
838	-33.1445221956073\\
1295	-35.0614142805904\\
2000	-36.9778091212037\\
};
\addlegendentry{\small MSE$\pg\widehat{\Pib}^t-\widehat{\Pib}^{GCWE}\pd$}

\addplot [color=black]
  table[row sep=crcr]{%
40	-20.0361868580188\\
62	-21.9395038397217\\
95	-23.7928229976277\\
147	-25.6887602922209\\
228	-27.5949354147437\\
352	-29.4810135795205\\
543	-31.3635852406276\\
838	-33.2480271310419\\
1295	-35.1382846289119\\
2000	-37.025886901379\\
};
\addlegendentry{\small $\mathcal{T}\pg\widehat{\Pib}^t-\widehat{\Pib}^{GCWE}\pd$}

\end{axis}
\end{tikzpicture}%
	\end{center}
	\caption{\label{fig6} Empirical and theoretical mean squared error on projector: Results for the  standard regime $\pg\widehat{\Pib}^t-{\Pib}\pd$ and for the Gaussian equivalent $\pg\widehat{\Pib}^t-\widehat{\Pib}^{\rm{GCWE}}\pd$; $t$-distributed data with $p = 20$, $r=5$, $d = 3$.
	}
\end{figure}
As detailed in Section \ref{sec:eigen}, this is expected since the eigenvector are scale-invariant functions of the scatter matrix. For the Student's $M$-estimator the first scale factor $\sigma_1$ is much smaller that $\sigma_2$, especially when the data dimension grows, and the approximation in this case is even stronger. 

%Figure \ref{fig5} displays the Frobenius norm of the psudo-covariance matrix of the first eigenvector. One can notice that both the norm of the pseudo-covariance matrix in the standard regime and the one for the Gaussian equivalent tend to zero. Unsurprisingly, the norm for the Gaussian equivalent goes to zero much faster.
 
Figure \ref{fig6} presents the MSE for the projector defined by \eqref{proj-M}. The data dimension $p$ is equal to 20, while the rank $r$ of $\Sig_r$
is set to 5. Parameter $\gamma^2$ is set to 1 and $\Lamb_r$ is designed such that $\text{min}\pg\text{diag}\pg\Lamb_r\pd\pd\gg\gamma^2$. The figure validates the theoretical results proposed in Theorems \ref{thm-3} and \ref{thm-4} and leads us to the same conclusions as previously.
%\begin{figure}[!h]
%%\setlength\fwidth{0.38\textwidth}
%\begin{center}
%\input{lranmf.tex}
%\end{center}
%\caption{Empirical and Theoretical MSE on lranmf; $M = 10$, $d = 2$
%}
%\label{fig4}
%\end{figure}

\subsubsection{SNR Loss}
Let us consider the STAP problem with the factor model introduced previously and the optimal filter $\wb_{opt}$ \cite{ward1994space} given by
\begin{equation}
	\wb_{opt}=\Sig^{-1}\pb
\end{equation}
where $\pb$ is the known STAP steering vector. In the low-rank clutter case an alternative is to use the low-rank STAP filter $\wb_{R}$ \cite{kirsteins1994adaptive,haimovich1997asymptotic} defined as
\begin{equation}
	\wb_{r}=\Pib^{\bot}\pb.
\end{equation}
In practice, in order to use the STAP filter one has to estimate the covariance matrix $\Sig$ and the projector $\Pib^{\bot}$ from the secondary data $\zb_i \sim \mathcal{CES}(\0b, \Sig, \gb_{\zb})$, which is usually done with an $M$-estimator. 

\begin{figure}[!h]
	\begin{center}
\begin{tikzpicture}[scale=0.8]
\pgfplotsset{every axis legend/.append style={fill=white,cells={anchor=west},at={(0.65,0.3)},anchor=north west}}
\begin{axis}[%
%width=0.951\fwidth,
%height=0.75\fwidth,
%at={(0\fwidth,0\fwidth)},
scale only axis,
xmode=log,
xmin=40,
xmax=2000,
xminorticks=true,
xlabel style={font=\color{white!15!black}},
xlabel={$n$},
ymin=-1.12,
ymax=0,
ylabel style={font=\color{white!15!black}},
ylabel={Mean Squared Error (dB)},
axis background/.style={fill=white},
title style={font=\bfseries},
title={},
xmajorgrids,
xminorgrids,
ymajorgrids,
legend style={legend cell align=left, align=left, draw=white!15!black}
]

\addplot [color=blue, mark=asterisk]
  table[row sep=crcr]{%
40	-0.656374850248736\\
62	-0.42245099194318\\
95	-0.271099634606838\\
147	-0.171790499196542\\
228	-0.11102865503619\\
352	-0.0716148512953629\\
543	-0.0461874583333393\\
838	-0.0301926367754704\\
1295	-0.0194894474084971\\
2000	-0.0125711768687595\\
};
\addlegendentry{SNR-ST}

\addplot [color=red, mark=asterisk, mark options={solid, red}]
  table[row sep=crcr]{%
40	-0.607107927216374\\
62	-0.397985293239543\\
95	-0.25572517929677\\
147	-0.164967573331578\\
228	-0.106437451885342\\
352	-0.0682242288084284\\
543	-0.0440436464569021\\
838	-0.0288437218633477\\
1295	-0.0186014653195995\\
2000	-0.0120039613798225\\
};
\addlegendentry{SNR-GCWE}

\addplot [color=black,mark=asterisk]
  table[row sep=crcr]{%
40	-0.579919469776867\\
62	-0.365168338257625\\
95	-0.234810958495229\\
147	-0.150289903651196\\
228	-0.096299839522931\\
352	-0.0621318868725728\\
543	-0.0401755392245776\\
838	-0.025990172234889\\
1295	-0.0168005811802167\\
2000	-0.0108709564121414\\
};
\addlegendentry{Eq. \eqref{thsnr}}

\addplot [color=green, mark=asterisk]
table[row sep=crcr]{%
	40	-1.11421432677155\\
	62	-0.888775716665405\\
	95	-0.704036158584843\\
	147	-0.553220341053622\\
	228	-0.438635804691227\\
	352	-0.350135352542428\\
	543	-0.283147425038216\\
	838	-0.22274656434559\\
	1295	-0.178750946905833\\
	2000	-0.143517948491631\\
};
\addlegendentry{SNR-SCM}

\end{axis}
\end{tikzpicture}%
	\end{center}
	\caption{\label{fig7} Empirical SNR Loss obtained with the Student's $M$-estimator (ST-SNR), GCWE (GCWE-SNR) and SCM (SCM-SNR) versus the theoretical result given by Eq. \eqref{thsnr}; $t$-distributed data with $p = 20$, $r=5$, $d = 3$.	}
\end{figure}
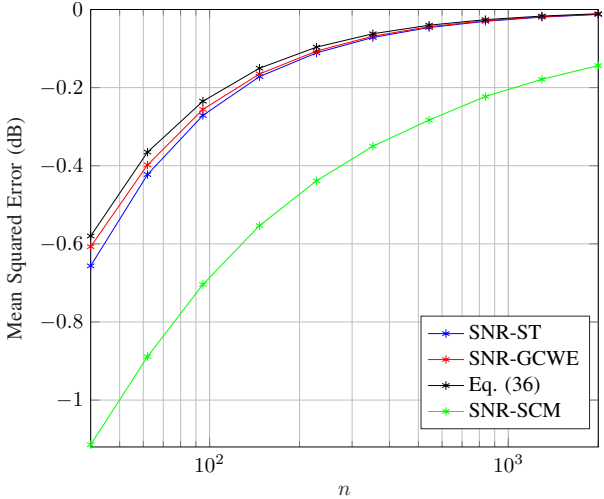

We are then interested in computing the SNR Loss $\rho$ given by 
\begin{equation}
	\rho=\frac{SNR_{out}}{SNR_{max}}=\frac{|\widehat{\wb}^H\pb|^2}{\pg\widehat{\wb}^H\Sig\widehat\wb\pd\pg\pb^H\Sig\pb\pd}
\end{equation}
or equivalently
\begin{equation}
	\rho=\gamma^2\frac{\pg\pb^H\widehat{\Pib}^{\bot}\pb\pd^2}{\pb^H\widehat{\Pib}^{\bot}\Sig\widehat{\Pib}^{\bot}\pb}.
\end{equation}
In \cite{haimovich1997asymptotic} it has been shown that when the data are Gaussian-distributed and the covariance matrix estimated using the SCM, the theoretical SNR Loss is given by 
\begin{equation}
\label{thsnr}
	\E\cg\rho\cd=1-\frac{r}{n}.
\end{equation}

Figure \ref{fig7} draws a comparison between the values of empirical mean of SNR Loss obtained with the projector estimate based on $\widehat{\Sig}_t$ with $t$-distributed secondary data (SNR-ST), empirical mean of SNR Loss computed with the corresponding GCWE (SNR-GCWE) which theoretical expectation is given in Eq. \eqref{thsnr}. One can notice that the value of SNR-ST is very close to the one SNR-GCWE, as anticipated, which supports the idea to approximate the behavior of SNR-ST with the one of SNR-GCWE \cite{haimovich1997asymptotic} when necessary. The green curve presents empirical mean of SNR Loss based on the SCM computed with $t$-distributed data showing the importance of $M$-estimators in CES context.

\section{Conclusion}
\label{sec:conclu}

This paper has analysed the asymptotic distribution of the EVD as well as the one of the principal subspace of scatter $M$-estimators. The results in the standard asymptotic regime have been derived. Then, relying on \cite{draskovic2018new_insights} the results have been extended, giving the convergence towards EVD of GCWE. The derived moments of second order appear to be much smaller in this case than in standard regime, offering a better approximation of the elements behavior. The applications of the theoretical results on SNR Loss and biased Intrinsic CLRB have been illustrated. We came up with the same conclusion that is that the behavior of EVD parameters are much better explained with the one of GCWE EVD parameters than with their standard asymptotic Normal distribution. The great benefit of these results is that one can use $M$-estimators to compute scatter matrix estimator and obtain a more precise  estimation of EVD in CES framework, while leaning on the theoretical results obtained for the corresponding GCWE. Importantly, these results can be easily applied to a wide scope of problems.

\appendices
\section*{Appendices}
To prove all theorems we will use the basic results obtained in the following theorem. 
\begin{lemma}
\label{thm-6}
Let $\ag\widehat{\zb}\ad$ be a sequence of complex random vectors $\widehat{\zb}$ and $\zb$ a compatible fixed vector. Assume that $\sqrt{N}\pg\widehat{\zb}-\zb\pd\overset{d}{\to}\mathcal{GCN}\pg\0b,\Vb,\Wb\pd$. Let $\xi\pg\yb\pd$ be a vector function of a vector $\yb$ with first and a second derivatives existing in a neighbourhood of $\yb=\zb$. Then
\begin{equation}
\label{delta}
\sqrt{N}\pg\xi\pg\widehat{\yb}\pd-\xi\pg\yb\pd\pd\overset{d}{\to}\mathcal{GCN}\pg\0b,\Db\Vb\Db^H,\Db\Wb\Db^T\pd 
\end{equation}
where 
\begin{equation}
\left.\Db=\frac{d\pg\xi\pg\yb\pd\pd}{d\yb}\right|_{\yb=\zb}
\end{equation}
is a matrix derivative.
\end{lemma}

\section{Proof of Theorem \ref{thm-1}}
\label{app1}
\begin{proof}
To derive the derivatives of $\phb$ and $\thb_j$ with respect to $\vec\pg\Mb\pd$ at the point $\Mb=\Sig$ we differentiate $\Mb\thb_j=\phi_j\thb_j$ 
\begin{equation}
\label{deriv}
d\Mb\ub_j+\Sig d\thb_j=d\phi_j\ub_j+\lambda_jd\thb_j.
\end{equation}
Multiplying each side of the last equation by $\ub^H_j$, one has
\begin{equation*}
d\phi_j=\ub^H_j\pg d\Mb\pd\ub_j
\end{equation*}
since $\ub^H_j\Sig=\lambda_j\ub^H_j$ and $\ub^H\ub=1$. Thus, 
\begin{equation*}
\left.\frac{d\phi_j}{d\vec\pg\Mb\pd}\right|_{\Mb=\Sig}=\ub^T_j\otimes\ub^H_j.
\end{equation*} 
If $\phb=(\phi_1,\hdots,\phi_p)$, then one has
\begin{equation*}
\left.\frac{d\phb_j}{d\vec\pg\Mb\pd}\right|_{\Mb=\Sig}=\Eb^T\pg\Ub^T\otimes\Ub^H\pd
\end{equation*}
with $\Eb=(\eb_1\otimes\eb_1\hdots\eb_p\otimes\eb_p)$ where $\eb_j$, $j=1,\hdots,p$ are unit vectors. Further, combining the statement given in Lemma \ref{thm-6} with Eq.\eqref{asymp-MC}, one obtains 
\begin{eqnarray}
\label{lamb}
&&\phantom{+}\Eb^T\pg\Ub^T\otimes\Ub^H\pd\pg\vartheta_1\pg\Sig^T_{\sigma}\otimes \Sig_{\sigma}\pd\pd\pg\Ub^*\otimes\Ub\pd\Eb \notag \\
&&+\Eb^T\pg\Ub^T\otimes\Ub^H\pd\vartheta_2\vec\pg\Sig_{\sigma}\pd\vec\pg\Sig_{\sigma}\pd^H \pg\Ub^*\otimes\Ub\pd\Eb \notag \\
&&=\vartheta_1\Eb^T\pg\Lamb^T\otimes\Lamb\pd\Eb+\vartheta_2\Eb^T\pg\vec\pg\Lamb\pd\vec\pg \Lamb\pd^H\pd\Eb \notag \\
&&=\vartheta_1\Lamb^2+\vartheta_2\lamb\lamb^T. \notag
\end{eqnarray}
Note that since the eigenvalues are real one obtains the same result using the expression for the pseudo-covariance matrix.

In order to obtain the results for eigenvectors, we will multiply Eq. \eqref{deriv} by $\ub^H_k$, $k\neq j$. Thus, one obtains
\begin{equation*}
\ub^H_k\pg d\Mb\pd\ub_j=\pg\lambda_j-\lambda_k\pd\ub^H_k d\thb_j
\end{equation*} 
as $\ub^H_k\ub_j=0$. Following the same steps as in \cite{kollo1993asymptotics} (done for the real case), it is easy to show that
\begin{eqnarray*}
d\thb_j=\sum_{j\neq k}\pg\lambda_j-\lambda_k\pd^{-1}\ub_k\ub_k^H\pg d\Mb\pd\ub_j+\ub_j\ub_j^H d\thb_j.
\end{eqnarray*}
In fact, the last element in the previous equality is omitted in the real case since $\ub_j^T d\thb_j=0$ (from $\thb_j^T\thb_j=1$). However, in the complex case $\ub_j^H d\thb_j\neq 0$, as from $\thb_j^H\thb_j=1$ one has $\ub_j^H d\thb_j+d\thb_j^H\ub_j=0$ and it is obvious that $\ub_j^H d\thb_j\neq d\thb_j^H\ub_j$. In some works, the authors use the different normalization for eigenvectors $\ub_j^H\thb_j=1$ which directly implies $\ub_j^Hd\thb_j=0$ and in those circumstances the results correspond to the ones in the real case. In the general (more common) case, one obtains
\begin{equation*}
\pg\Iden-\ub_j\ub_j^H \pd d\thb_j=\pg\ub_j^T\otimes\Ub\pg\lambda_j\Iden-\Lamb\pd^+\Ub^H\pd d\Mb,
\end{equation*}
which actually gives the projection of the derivative onto the subspace orthogonal to the one of the eigenvector. Now, employing Eq. \eqref{delta} with the previous derivatives and since 
\begin{eqnarray}
\pg\ub_j^T\otimes\Ub\pg\lambda_j\Iden-\Lamb\pd^+\Ub^H\pd\Kb=\Ub\pg\lambda_j\Iden-\Lamb\pd^+\Ub^H\otimes\ub_j^T \notag \\
\pg\lambda_j\Iden-\Lamb\pd^+\eb_j=\0b \notag \\
\cg \ub_j^T\otimes\Ub(\lambda_j\Iden-\Lamb)^+\Ub^H\cd\vec\pg\Sig\pd=\0b \notag
\end{eqnarray}
one obtains the final results. Note that $\mathcal{GCN}$ becomes $\mathcal{CN}$ since the pseudo-covariance matrix is equal to zero.
\end{proof}

\section{Proof of Theorem \ref{thm-2}}
\label{app2}
\begin{proof}
Rewriting the left-hand side of Eq. \eqref{eig-MC}
\begin{eqnarray}
\sqrt{n}\pg\sigma\widehat{\lamb}^{M}-\widehat{\lamb}^{\rm{GCWE}}\pd=\sqrt{n}\pg\sigma \widehat{\lamb}^{M}-\lamb-\widehat{\lamb}^{\rm{GCWE}}+\lamb\pd=\notag \\
\sqrt{n}\pg\pg\sigma\widehat{\lamb}^{M}-\lamb\pd-\pg\widehat{\lamb}^{\rm{GCWE}}-\lamb\pd\pd.\notag 
\end{eqnarray}
Then 
\begin{eqnarray}
\text{var}_n\pg\sigma\widehat{\lamb}^{M}-\widehat{\lamb}^{\rm{GCWE}}\pd=\notag\\
\E\cg n\pg\sigma\widehat{\lamb}^{M}-\widehat{\lamb}^{\rm{GCWE}}\pd\pg\sigma\widehat{\lamb}^{M}-\widehat{\lamb}^{\rm{GCWE}}\pd^T\cd \notag \\
=\text{var}_n\pg\sigma\widehat{\lamb}^{M}\pd-2\text{cov}_n\pg\widehat{\lamb}^{M},\sigma\widehat{\lamb}^{\rm{GCWE}}\pd+\text{var}_n\pg\widehat{\lamb}^{\rm{GCWE}}\pd \notag.
\end{eqnarray}
Since from \eqref{eig-MC} one has
\begin{eqnarray}
&&\text{var}_n\pg\sigma\widehat{\lamb}^{M}\pd\xrightarrow [n\to+\infty]{}\vartheta_1\Lamb^2+\vartheta_2\lamb\lamb^T \quad \text{and}\notag \\
&&\text{var}_n\pg\widehat{\lamb}^{\rm{GCWE}}\pd\xrightarrow [n\to+\infty]{}\Lamb^2, \notag
\end{eqnarray}
it remains only to derive the expression for 
\begin{equation*}
\text{cov}_n\pg \sigma\widehat{\lamb}^{M},\widehat{\lamb}^{\rm{GCWE}}\pd=\E\cg n\pg\sigma\widehat{\lamb}^{M}-\lamb\pd\pg\widehat{\lamb}^{\rm{GCWE}}-\lamb\pd^T\cd.
\end{equation*}
Using the Delta method, one can show that
\begin{eqnarray}
\label{lambcov}
\text{cov}_n\pg \sigma\widehat{\lamb}^{M},\widehat{\lamb}^{\rm{GCWE}}\pd\to\left.\frac{\sigma d\phb}{d\vec\pg\Mb\pd}\right|_{\Mb=\Sig}\Qb\left.\frac{d\phb}{d\vec\pg\Mb\pd}\right|_{\Mb=\Sig}^T  \notag
\end{eqnarray}
where $\Qb$ is the asymptotic pseudo-covariance matrix of $\widehat{\Sig}$.
This matrix is equal to 
\begin{equation}
\Qb=\gamma_1 \pg\Sig^T_{\sigma}\otimes \Sig_{\sigma}\pd\Kb+\gamma_2\vec\pg\Sig_{\sigma}\pd\vec\pg\Sig_{\sigma}\pd^T
\end{equation}
as given in  ref. Repeating the same steps as in Eqs. \eqref{lamb}, one shows that the right-hand side of the right-hand side of Eq. \eqref{lambcov} becomes
\begin{equation*}
\gamma_1\Lamb^2+\gamma_2\lamb\lamb^T
\end{equation*}
which, since $\sigma_1=\vartheta_1-2\gamma_1+1$ and $\sigma_2=\vartheta_2-2\gamma_2$, leads to the final results.

The results for the eigenvectors can be obtained following the same procedure as for the eigenvalues.
\end{proof}

\section{Proof of Theorem \ref{thm-3}}
\label{app3}

\begin{proof}
If we define the pseudo-inverse of $\Sig_r$ as
\begin{equation}
\label{psinv}
\Phb=\Ub_r\Lamb^{-1}_r\Ub_r^H
\end{equation} 
one has from ref that
\begin{equation*}
\widehat{\Pib}_r=\Pib_r+\delta\Pib_r+\hdots+\delta^i\Pib_r+\hdots
\end{equation*}
where
\begin{eqnarray}
\label{projser}
&&\delta\Pib_r=\Pib_r^{\bot}\Delta\Sig\Phb+\Phb\Delta\Sig\Pib_r^{\bot} \notag \\
&&\delta^i\Pib_r=-\Pib_r^{\bot}\pg\delta^{i-1}\Pib\pd\Delta\Sig\Phb+\Pib_r^{\bot}\pg\delta^{i-1}\Pib\pd\Delta\Sig\Phb \notag
\end{eqnarray}
with $\Delta\Sig=\widehat{\Sig}-\Sig$.

In the asymptotic regime, when $n\to\infty$ we can write
\begin{equation*}
\widehat{\Pib}_r=\Pib_r+\delta\Pib_r
\end{equation*}
since $\Delta\Sig$ is close to zero. Hence, taking a vec of the $\widehat{\Pib}_r-\Pib_r=\delta\Pib_r$, one gets
\begin{equation*}
\vec\pg\widehat{\Pib}_r-\Pib_r\pd=\Fb\vec\pg\widehat{\Sig}-\Sig\pd
\end{equation*}
with
\begin{equation*}
\Fb=\pg\Phb^T\otimes\Pib_r^{\bot}+\pg\Pib_r^{\bot}\pd^T\otimes\Phb\pd.
\end{equation*}
It is now obvious that the covariance ($\resp$pseudo-covariance) matrix of $\sqrt{n}\pg\Pib_r^M-\Pib_r\pd$ is equal to $\Fb\Cb\Fb^H$ ($\resp$ $\Fb\Pb\Fb^T$) where $\Cb$ and $\Pb$ are given in Eqs. \eqref{asymp-MC}. Further
\begin{eqnarray}
&&\Fb\Cb=\pg\Phb^T\otimes\Pib_r^{\bot}+\pg\Pib_r^{\bot}\pd^T\otimes\Phb\pd\pg\Sig^T\otimes\Sig\pd \notag \\
&&+\pg\Phb^T\otimes\Pib_r^{\bot}+\pg\Pib_r^{\bot}\pd^T\otimes\Phb\pd\vec\pg\Sig\pd\vec\pg\Sig\pd^H \notag \\
&&=\pg\Phb^T\Sig^T\otimes\Pib_r^{\bot}\Sig+\pg\Pib_r^{\bot}\pd^T\Sig^T\otimes\Phb\Sig\pd \notag
\end{eqnarray}
as $\pg\Phb^T\otimes\Pib_r^{\bot}+\pg\Pib_r^{\bot}\pd^T\otimes\Phb\pd\vec\pg\Sig\pd=\0b$ using $\pg\Tb^T\otimes\Rb\pd\vec\pg\Sb\pd=\vec\pg\Rb\Sb\Tb\pd$ and $\Pib_r^{\bot}\Sig\Phb=\Phb\Sig\Pib_r^{\bot}=\0b$.
Finally, after the postmultiplication by $\Fb^H$ and since
\begin{eqnarray}
&&\Sig=\Sig^H\neq\Sig^T \notag \\
&&\Phb=\Phb^H\neq\Phb^T \notag \\
&&\Pib_r^{\bot}=\pg\Pib_r^{\bot}\pd^H\neq\pg\Pib_r^{\bot}\pd^T \notag 
\end{eqnarray} 
one obtains
\begin{eqnarray}
\Fb\Cb\Fb^H=\pg\pg\Phb\Sig\Phb\pd^T\otimes\Pib_r^{\bot}\Sig\Pib_r^{\bot}+\pg\Pib_r^{\bot}\Sig\Pib_r^{\bot}\pd^T\otimes\Phb\Sig\Phb\pd \notag
\end{eqnarray}
which with $\Phb$ given by Eq. \eqref{psinv} and $\Sig=\Ub_r\Lamb_r\Ub_r^H+\gamma^2\Iden_p$ yields the final result.

Analogously, one can derive the results for the pseudo-covariance using the equality $\Kb\pg\Ab\otimes\Bb\pd=\pg\Bb\otimes\Ab\pd\Kb$.
\end{proof}

\section{Proof of Theorem \ref{thm-4}}
\label{app4}

Following the steps from Appendix \ref{app2} and using the results of Theorem  \ref{thm-3} one gets the results of the theorem.

\bibliographystyle{IEEEtran} 
\bibliography{biblio_new}

\end{document}